\documentclass[preprint,showpacs]{revtex4}
\usepackage{graphicx,amsmath}

\begin{document}
\title{Resultant pressure distribution pattern along the basilar membrane in the
spiral shaped cochlea}

\author{Yong Zhang}
\email{xyzhang@phya.yonsei.ac.kr}
\author{Chul Koo Kim}
\email{ckkim@yonsei.ac.kr}
\affiliation{Institute of Physics and Applied Physics, Yonsei
University, Seoul 120-749, Korea}
\author{Kong-Ju-Bock Lee}
\email{kjblee@ewha.ac.kr}
\affiliation{Department of Physics, Ewha Womans University, Seoul 120-750,
Korea}
\affiliation{School of Physics, Korea Institute for Advanced Study, Seoul
130-722, Korea}
\author{Youngah Park}
\email{youngah@mju.ac.kr}
\affiliation{Department of Physics, Myongji University, Yongin 449-728,
Korea}
\affiliation{School of Physics, Korea Institute for Advanced Study, Seoul
130-722, Korea}

\begin{abstract}
Cochlea is an important auditory organ in the inner ear. In most
mammals, it is coiled as a spiral. Whether this specific shape influences
hearing is still an open problem. By employing a three dimensional
fluid model of the cochlea with an idealized geometry, the influence of 
the spiral geometry of the cochlea is examined. We obtain solutions of the 
model through a conformal transformation in a long-wave approximation.
Our results show that the net pressure acting
on the basilar membrane is not uniform along its spanwise direction.
Also, it is shown that the location of the maximum of the spanwise 
pressure difference in the
axial direction has a mode dependence. In the simplest pattern, the
present result is consistent with the previous theory based on the
WKB-like approximation [D. Manoussaki, {\it et al.,} Phys. Rev. Lett. {\bf
96}, 088701~(2006)]. In this mode, the pressure difference in the spanwise
direction is a monotonic function of the distance from the apex and the
normal velocity across the channel width is zero. Thus in the lowest order
approximation, we can neglect the existance of the Reissner's membrane in
the upper channel. However, higher responsive modes show
different behavior and, thus, the real maximum is expected to be
located not exactly at the apex, but at a position determined by the
spiral geometry of the cochlea and the width of the cochlear duct. In
these modes, the spanwise normal velocities are not zero. Thus, it
indicates that one should take into account of the detailed geometry of the
cochlear duct for a more quantitative result.
The present result clearly demonstrates that not only the spiral geometry,
but also the geometry of the cochlear duct play decisive roles in distributing
the wave energy.
\end{abstract}
\maketitle

\section{Introduction}
The primary functions of the mammalian cochlea are to
detect and analyze the acoustic signals in terms of their intensity and
frequency contents. It is able to probe and analyze
sound waves over wide ranges of frequency and intensity, e.g., for
human being, over twelve orders in intensity and over
three orders in frequency ($20$Hz -- $20$kHz)~\cite{RMP_v12_47}.
Some mammals are able to perceive frequencies as high as
$100$kHz~\cite{PhysiolRev_v81_1305}. Remarkable capabilities appearing
in hearing are essentially governed by both passive mechanical and
active biophysical procedures in the cochlea. The physics of hearing is
pioneered by von Helmholtz in 1880s~\cite{helmholtz}. He
proposed that different regions along the longitudinal length of
the cochlea vibrate independently, working as a series of oscillators
driven by the pressure of the fluid filled in the cavity of the cochlea.
The coupling among the successive parts of the cochlea is thought to be
mainly determined through the fluid, instead of the cochlea itself.
Benefitted from the achievements in both the experimental techniques and
the theoretical methods, auditory physics has been developing continuously
from the time of von Helmholtz and expanding on ever-widening problems
~\cite{PhysiolRev_v81_1305,Dallos,Gummer,HC}.

Observations revealed that the cochleae are coiled in almost all mammals,
monotremes being the exceptions. This auditory apparatus is a fully
fluid-filled system of canals contained inside the petrous bone. It
suggests that detailed calculation of the fluid motion and the pressure
distribution within the cochlea should be crucial for understanding
of hearing in any mechanical models of the cochlea. Fluid in the
cochlea is a lymph. Physically, it can be treated as incompressible
and inviscid as a reasonable approximation~\cite{CON_v7_480}. The cavity
of the cochlea is partitioned by the basilar membrane (BM) into two
channels which are connected through the
helicotrema at the apex of the cochlea. At the other end, the base,
the channels are sealed by two movable membranes
working as pistons. These two movable membranes are labeled as the oval
and the round windows, respectively. The channel abutting to the oval
window is named as scala vestibuli (upper channel). Another channel
abutting to the round window is known as scala tympani (lower
channel). The upper channel is further separated by
the Reissner's membrane (RM) into two subducts. However,
the RM is usually ignored in the mechanical models for the cochlea because
of its extreme flexibility, when the axial flow behavior is the main focus. 
As a consequence, the structure of the cochlea is
typically modeled by two channels with the BM as their interface in most
mechanical studies. However, when the spanwise pressure distribution is
included in the study, the RM affects the flow characteristics and, thus,
cannot be neglected. We will address this problem below.

The stapes is attached on the oval window and forms the boundary
between the middle and the inner ear.
Incoming sound waves are transmitted into the fluid in the
channels of the cochlea through the vibration of the stapes. Nearly
entire mechanical sound energy exerted by the stapes is used to
produce the fluid movement. The sound pressure transmitted to the cochlear
fluid interacts mechanically with the BM and vibrates it into
oscillation in the form of a traveling wave that propagates along
the BM from the base to the apex. The traveling wave on the BM
grows as it travels, reaching a maximum at a position known as
the characteristic place determined by the frequency
of the stimulus, and then declines rapidly. The characteristic
behavior along the longitudinal cochlea is described by a
place-frequency map; high frequencies are resonant near the basal end,
whereas low frequencies are resonant near the apical end~\cite{Dallos}.

The first direct observation of the responses of the BM to sounds through
experiment was achieved by von B\'{e}k\'{e}sey~\cite{von}. He
measured the BM movement of the cadaver and found that the mechanical
response is a traveling wave propagating from the base to the apex.
Subsequent in-vivo measurements revealed that the responses of the BM
are sharply tuned and very sensitive at low sound-pressure levels.
Hence it is evident that hearing involves an active process. However,
a passive mechanism also plays an essential role in hearing and also 
in understanding the auditory phenomena.

Although the spiral coiling is one of the most distinct characters of
the cochlea, it has been neglected in many mechanical
models~\cite{Gummer}. Therefore, it is one of the prime interests
to ask whether the shape of the cochlea influences hearing
or just facilitates the packing of the long BM into a small space. If
the answer is the latter, then the cochlea can be safely modeled as
a straightly shaped duct to avoid the complexity of the spiral shape.
Due to the mathematical difficulties in dealing with the shape
of the cochlea, this problem has not been fully answered yet. Earlier
studies mostly concluded that the shape of the cochlea has
little effect on the response of the BM along the center
line~\cite{von,JASA_v64_1048,JASA_v74_95,JASA_v77_1849}. However,
recently, Manoussaki {\it et. al.} employed a WKB-like and linearly
approximated solution to examine the effect of the curvature of the cochlea
on the redistribution of the wave energy density. According to their linear
expansion, it was concluded that
the spiral geometry of the cochlea does play an essential role in
creating a tilt of the BM between the outer and the inner wall, which is
believed to be important to produce the vibration of the hair cells.
The calculated tilt is shown to be proportional to the curvature of the
center line of the cochlea and, thus, the maximum of the tilt is
predicted to be always at the apex~\cite{PRL_v96_088701}. In the
present paper, we employ a three dimensional fluid model without the
assumption of the WKB-like solution. By employing a conformal
transformation and a long-wave approximation, we explicitly obtain the 
solutions of the fluid model which satisfies the boundary condition. 

Our analytic solution shows that the Manoussaki {\it et al}'s 
solution~\cite{PRL_v96_088701} is naturally included as the lowest component 
of the general solutions, and has no flux across the channel width. The 
general solution shows that the resultant pressure acting on the BM is not 
uniform across the channel width. It is shown that different responsive modes 
show different patterns of the spanwise 
pressure tilt along the longitudinal length of the BM. Especially, for higher 
modes which allow the flux across the channel width, the magnitude of the 
pressure tilt is not always a monotonic function of the curvature as 
predicted by Manoussaki {\it et al.}~\cite{PRL_v96_088701}. Since the real
fluid oscillation is a linear superposition of several modes, we
expect that the the maximum of the resultant pressure tilt is not located
exactly at the apex as predicted by Manoussaki {\it et
al.}~\cite{PRL_v96_088701}, but at a position determined by the
input wave and the geometry of the cochlea.

The arrangement of this paper is as follows. In Sec. II, we give the
fluid model of the cochlea and calculate the velocity potential and the
flow of the fluid in the duct of the cochlea. Sec. III discusses the
pressure in the fluid using the Navier-Stokes equation. Sec. IV
gives a discussion on the results. Sec. V concludes the paper.

\section{Mechanical Model}
Because most mammalian cochleae are coiled into a spiral, the
cochlea is approximately modeled as a logarithmic spiral duct with a
rectangular cross section and the RM is approximately considered as
the center line in the upper channel as shown in Fig.~1, although the
detailed geometry is much more complicated. We also assume that it stays 
in plane in the present model although the center line of the cochlea 
coils out-of-plane in reality, The BM partitions the duct into two channels, 
the upper and the lower channels. The height and the width of the cochlea 
are assumed uniform along its longitudinal length. In what follows, we 
mainly discuss the response of the fluid in the outer part of the upper 
channel, between the RM and the outer wall using our idealized geometry. 
The result can be extended to the lower channel by considering the 
connectivity of the helicotrema and the symmetry as elaborated in the 
references~\cite{PRL_v96_088701,KernThesis}. Cochlea fluid is 
approximated as incompressible and irrotational, i.e., $\mathbf{\nabla} 
\cdot \mathbf{U} = 0$ and $\mathbf{\nabla} \times \mathbf{U} =0$, where
$\mathbf{U}$ is the velocity of the fluid. Hence the velocity of the
fluid in the cochlea can be described in terms of the velocity
potential, $\mathbf{U} = \mathbf{\nabla} \Phi$. Thus, the velocity
potential $\Phi$ satisfies the Laplace equation
\begin{eqnarray}
\nabla^2 \Phi = 0. \label{vp_lp}
\end{eqnarray}
In the cylindrical polar coordinate system,
\begin{eqnarray}
\nabla^2 = \frac{\partial^2}{\partial R^2} +
\frac{1}{R}\frac{\partial} {\partial R} +
\frac{1}{R^2}\frac{\partial^2}{\partial \Theta^2} +
\frac{\partial^2}{\partial \zeta^2},
\end{eqnarray}
where $R$ and $\Theta$ are the coordinates in the plane of the BM; $\zeta$
is the coordinate perpendicular to the BM. At the apex, the polar angle
is zero and at the spiral axis, $R=0$. Recently Manoussaki {\it et al.}
studied the cochlea curvature effect on the radial shearing in the BM
region~\cite{PRL_v96_088701}. Instead of solving the Laplace equation,
Eq.~(\ref{vp_lp}) directly, they assumed a WKB-like propagating wave
solution,
\begin{eqnarray}
\Phi = \Phi_0(R,\Theta,\zeta)\exp \left[i\omega t - i\int_0^{\Theta} K(\Theta)
R_m(\Theta)d\Theta\right], \label{wkb}
\end{eqnarray}
where $\Phi_0$ is a slowly varying function of the angular coordinate
$\Theta$, $R_m$ the distance to the spiral duct midline from the spiral
axis, $K$ the axial wave number, and $\omega$ the input frequency. They
applied this solution to the continuity and interface equations of the BM
motion to obtain the BM displacement. In the first order approximation,
they found that the displacement of the BM-fluid interface gets tilted
across the width of the partition and the spanwise amplitude slope varies as
$1/R_m$. Thus, they predicted that the maximum of the tilt to be always at
the apex in the long-wavelength limit and the spiral shape of the cochlea
plays an important role in distributing the wave energy density towards the
cochlea's outer wall.

However, it is believed that the pressure is transmitted through the fluid
and not through the BM itself. Therefore, it is necessary to study the
detailed fluid response to the applied pressure in order to obtain detailed
information on the pressure redistribution and the sound propagation. In
the present study, we solve the Laplace equation directly without the
assumption of the WKB-like plane wave solution. Since the above equation
does not contain any explicit time-dependence, solutions of this equation
cannot describe propagation of the waves along the
cochlea. However, it can give information on the spanwise distribution of
the pressure on the BM of the steady response. It has been demonstrated 
that at least to the first order approximation, the in-phase
assumption is a good approximation to discuss the spanwise
profile of the BM~\cite{JASA_v116_1025}.
Separating the variable, $\zeta$, from the rest two variables, we have
$\Phi(R,\Theta,\zeta,t) = \phi(R,\Theta) H(\zeta)T(t)$.
Eq.~(\ref{vp_lp}) is separated into two independent parts
\begin{eqnarray}
&& \frac{\partial^2 H(\zeta)}{\partial \zeta^2} = \mu^2H(\zeta),
\label{LP1} \\
&& \frac{\partial^2 \phi}{\partial R^2} + \frac{1}{R}\frac{\partial
\phi}{\partial R} + \frac{1}{R^2}\frac{\partial^2 \phi}{\partial
\Theta^2} = - \mu^2\phi, \label{LP2}
\end{eqnarray}
where $\mu$ is a separation constant that can be any constant.
Here, we note that if we apply the WKB-like solution, Eq.~(\ref{wkb}), to
Eq.~(\ref{LP2}), we obtain the same result as in Ref.~\cite{PRL_v96_088701}.
Now, we impose a boundary condition that the wall of the cochlea is
impermeable, i.e., the velocity of the fluid normal to the wall is
zero on the walls of the cochlea. However, because the RM is extremely flexible,
the velocity normal to the RM is not necessarily zero. Since the RM is
fixed on the BM, the pressure distribution near the intersection line
between the RM and the BM is expected to be much complicated. The
solution for such a case can only be obtained through very
complicated numerical calculations. Since, it is the aim of this paper
to investigate the curvature effect of the cochlea and improve from
the WKB-like solution, we neglect the edge effect and assume that the RM
behaves as a free bondary. Therefore, the
solution of the velocity potential in the vertical direction to the BM is
\begin{eqnarray}
H(\zeta) = \cosh\left[\mu(\zeta - h)\right], \label{vs}
\end{eqnarray}
where $h$ is the half height of the cochlea. For simplicity, we assume that
$\mu$ is sufficiently small, $\mu \ll 1$. Note that the small $\mu$
corresponds to a long-wave approximation~\cite{PRL_v96_088701,JFM_v106_149}.
We will discuss later that the small $\mu$ is also related to a weak
stimulation. Accordingly, Eq.~(\ref{LP2}) is approximately rewritten as
\begin{eqnarray}
\frac{\partial^2 \phi}{\partial R^2} + \frac{1}{R}\frac{\partial
\phi}{\partial R} + \frac{1}{R^2}\frac{\partial^2 \phi}{\partial
\Theta^2} = 0. \label{LP3}
\end{eqnarray}
The solution of Eq.~(\ref{LP3}) thus can be obtained as
\begin{eqnarray}
\phi(R,\Theta) &=& (A_l\cos l\Theta + B_l\sin
l\Theta)R^l,
\end{eqnarray}
where $l$ is any integer and $A_l(B_l)$ the constants to be
determined by the boundary conditions. $R^{-l}$ terms are neglected
to avoid diverging contributions at small $R$. In order to employ
techniques developed in the harmonic functions, the velocity
potential, $\phi$, is rewritten in a complex plane
\begin{eqnarray}
\phi = \left[A_l(z^l + \bar{z}^l) -i B_l(z^l - \bar{z}^l)
\right], \label{hs1}
\end{eqnarray}
where $z = R\exp(i\Theta)$ and $\bar{z}$ is the complex conjugate of $z$.
The solution should satisfy the boundary condition, impermeability of
the spiral walls. It thus requires to recombine these separated variables
together in an appropriate way, so that the spiral boundary
condition can be conveniently achieved. Conformal transformation is a
natural way to achieve it. We introduce a complex potential,
$f = \phi + i\varphi$. For $f$ being analytic,
$\phi$ and $\varphi$ must satisfy the
Riemann-Cauchy relation,
\begin{eqnarray}
\frac{\partial \phi}{\partial x}  = \frac{\partial \varphi}{\partial y},
\frac{\partial \phi}{\partial y}  = -\frac{\partial
\varphi}{\partial x}, \label{riemann}
\end{eqnarray}
where $x$ and $y$ are the real and the imaginary
parts of the complex variable, $z$, i.e., $z = x + iy$. It is easy to
verify that the imaginary part of the complex potential is the streamline of
the fluid, $\varphi = -[A_l(z^l + \bar{z}^l) + iB_l(z^l -
\bar{z}^l)]$. The complex potential is thus
\begin{eqnarray}
f = z^l\left[A_l - iB_l\right]. \label{CP1}
\end{eqnarray}
Now, we introduce a conformal transformation
\begin{eqnarray}
w = \rho_o\mathrm{e}^{-iz},
\end{eqnarray}
where $\rho_o$ is chosen to be the distance from
the origin of the spiral axis to the outer wall of the apex.
After the conformal transformation, the complex potential has the form
\begin{eqnarray}
f = \left(i\ln\frac{w}{\rho_o}\right)^l (A_l -
iB_l).
\end{eqnarray}
Using that $w = r\mathrm{exp}[i\theta]$ in the polar representation,
\begin{eqnarray}
f = \left[- \theta + i\ln \frac{r}{\rho_o}\right]^l (A_l -
iB_l). \label{CP_final}
\end{eqnarray}
The horizonal velocity potential $\phi(r,\theta)$ is the real part
of the complex potential~\cite{math}. Combining the horizonal and the
vertical parts of the velocity potential, one has
\begin{eqnarray}
\Phi(r,\theta,\zeta,t) = \phi(r,\theta)H(\zeta)T(t) = Re(f)
H(\zeta)T(t), \label{Phi}
\end{eqnarray}
where $Re(f)$ is the real part of the complex potential $f$.
Accordingly, the velocity potential is also a function of $l$. Different
$l$ corresponds to each different responsive mode.
To distinguish different modes, we introduce a suffix $l$ into
the two-dimensional and the three dimensional
velocity potential, velocity and pressure, i.e., $\phi_l$, $\Phi_l$,
$U_l$ and $P_l$, respectively. In general,
\begin{eqnarray}
\phi_l(r,\theta) &=& A_l\sum_{n=0}^{\lfloor l/2\rfloor}(-1)^{l-n}
{l \choose 2n}\left(\ln\frac{r}{\rho_o} \right)^{2n}\theta^{l-2n}
\nonumber \\
&+& B_l\sum_{m=1}^{\lfloor (l+1)/2\rfloor} (-1)^{l-m}{l \choose 2m-1}
 \left(\ln\frac{r}{\rho_o}\right)^{2m-1}\theta^{l-2m+1}, \label{gvp}
\end{eqnarray}
where $\lfloor x\rfloor$ represents the largest integer that is smaller
than or equal to $x$.
The above solution shows that the originally separated variables are
recombined through the conformal transformation in the process of
generating the spiral boundary.

The position vector of the logarithmic spiral is $\mathbf{r} =
\rho\mathrm{e}^{b\theta}\hat{e}_r$ in the polar coordinate system, where
$\rho$ is the distance from the spiral axis to the apical end of the
spiral when the polar angle is zero; $b$ is a parameter describing
the spiral. The covering area of the spiral increases with $b$
describing the compactness of the logarithmic spiral. On the outer
wall, $\mathbf{r}_o = \rho_o \mathrm{e}^{b\theta}\hat{e}_r$,
the boundary condition of the impermeability of the cochlea wall
requires that the normal component of the velocity should be zero;
\begin{eqnarray}
U_{l,n}(\mathbf{r}_o,\theta) = \mathbf{U}_l(\mathrm{r}_o,\theta) \cdot
 \hat{e}_n = 0,
\end{eqnarray}
where $\hat{e}_n$ is the unit vector normal to the spiral (see Appendix A).
This condition provides relations between $A_l$ and $B_l$ as follows for
$l=1,2,3$;
\begin{eqnarray}
B_1 = -bA_1, \label{a1b1}
\end{eqnarray}
\begin{eqnarray}
B_2 = \frac{2b}{b^2 - 1}A_2, \label{a2b2}
\end{eqnarray}
\begin{eqnarray}
B_3 = \frac{(3-b^2)b}{3b^2-1}A_3. \label{a3b3}
\end{eqnarray}
And the velocity potentials for the three lowest modes are
\begin{eqnarray}
\phi_1(r,\theta) = -A_1\left(\theta + b\ln\frac{r}{\rho_o}\right),
\label{po1}
\end{eqnarray}
\begin{eqnarray}
\phi_2(r,\theta) = A_2\left(\theta^2 - \ln^2\frac{r}{\rho_o} +
 \frac{4b}{1-b^2}\theta\ln\frac{r}{\rho_o}\right), \label{po2}
\end{eqnarray}
\begin{eqnarray}
\phi_3(r,\theta) &=& A_3\left[\theta\left(3\ln^2 \frac{r}{\rho_o} -
 \theta^2\right) \right. \nonumber \\
&-& \left. \frac{(3-b^2)b}{1-3b^2}\ln\frac{r}{\rho_o} \left(3\theta^2 -
 \ln^2\frac{r}{\rho_o}\right)\right]. \label{po3}
\end{eqnarray}
Since, we assumed that the RM is extremely flexible and, thus, behaves as a
free boundary, the
normal velocities of the inner side and the outer side are to be the same
on the RM. We explicitly get the solution of the velocity potential of the 
outer part of the upper channel. According to Eq.~(\ref{po1}), the normal
velocity is always zero in the first mode. Consequently, this solution can
be directly extended to the inner part of the cochlear duct. In this case,
the fluid only has angular flux and the RM can be neglected in the
mechanical model. However, for Eqs.~(\ref{po2}) and (\ref{po3}), the fluid
possesses normal flux. The normal velocity on the outer wall is zero but it
is nonzero on the RM. Thus, it is expected that the RM critically
influences fluid flow for the higher modes. In order to investigate
this point more closely, we carried out a similar calculation on the
inner part, from the inner wall to the RM. We find the normal velocities
of the inner side and the outer side on the RM are virtually the same. 
Thus, we belive that when the stimulation is weak so that no excessive
pressure can be build up at the RM and at the BM boundary, the assumption of
the free boundary for the RM can provide at least a qualitative picture for
the BM oscillation.

\section{Pressure in the fluid acting on the BM}
The pressure of the fluid is governed by the Navier-Stokes (N-S)
equation~\cite{Lighthill}. Viscosity of the lymph in the cochlear duct is
usually negligible in the studies of hearing~\cite{CON_v7_480}. The
inviscid N-S equation reads
\begin{eqnarray}
\rho_{mass} \frac{D\mathrm{U}}{Dt} = -\nabla P , \label{NS}
\end{eqnarray}
where $\rho_{mass} $ is the density of the fluid; $D/Dt =
\partial/\partial t + \mathbf{U}\cdot \nabla$; $P$ is the
pressure of the fluid. In the study of hearing, the
disturbance of the fluid in the cochlea is usually small. Therefore, the
pressure in the fluid can be obtained through the linearized N-S
equation~\cite{Lighthill}
\begin{eqnarray}
P = -\rho_{mass} \frac{\partial \Phi}{\partial t}.  \label{P1}
\end{eqnarray}

The pressure in different modes can thus be obtained through the linearized
N-S equation
\begin{eqnarray}
P_l = -i \rho_{mass}\omega \phi_l(r,\theta)
\cosh[\mu(\zeta-h)]T(t), \label{pressure}
\end{eqnarray}
where we assume an input frequency $\omega$. The fluid pressure right above
and below the BM are $P_{\mathrm{up}}$ and $P_{\mathrm{down}}$.
When the upper channel is driven alone, the cochlear duct is expected
to reduce the encountered large impedance through the lower channel. It
has been shown that due to the symmetric structure of the upper and the 
lower channels and the connectivity of the helicotrema, we obtain an
approximate relation $P_{\mathrm{up}} \approx
-P_{\mathrm{down}}$~\cite{PRL_v96_088701,KernThesis}.
Hence the net pressure in the mode $l$, which is the pressure
difference between the upper and the lower channels acting on the BM, is
\begin{eqnarray}
 P_l^{\mathrm{BM}}(r,\theta,t) = -2P_l(r,\theta,0,t), \label{dp}
\end{eqnarray}
where $P_l$ is the pressure in the upper duct in the mode $l$.
We define a dimensionless tilt of pressure on the BM, which can be directly
compared to the BM displacement tilt in Ref.~\cite{PRL_v96_088701}, as
\begin{eqnarray}
\Delta P_l^{\mathrm{BM}} \equiv \frac{P_l^{\mathrm{BM}}(r_o,\theta) -
 P_l^{\mathrm{BM}}(r_c,\theta)}{P_l^{\mathrm{BM}}(r_o,\theta_{\mathrm{
 base}})},
\end{eqnarray}
where $r_o(\theta)$ is the distance to the outer wall following the spiral
arc, $\theta_{\mathrm{base}} = 2K\pi$ with $K$ being the number of the spiral
turns, and $r_c(\theta)$ the distance to the center of the BM.

An arbitrary position vector on the BM can be expressed as $\mathbf{r} =
\mathbf{r}_o + \sigma \hat{e}_n$, where $\sigma$ is the normal distance
from the outer wall to the position. Hence
\begin{eqnarray}
\mathbf{r} = \left(\rho_o\mathrm{e}^{b\theta} - \frac{\sigma}{\sqrt{b^2+1}}
 \right)\hat{e}_r + \frac{b\sigma}{\sqrt{b^2+1}}\hat{e}_{\theta},
\end{eqnarray}
\begin{eqnarray}
r = |\mathbf{r}| =r_o g\left(\frac{\sigma}{R_o}\right),
\end{eqnarray}
\begin{eqnarray}
g \left(\frac{\sigma}{R_o}\right) = \sqrt{1-\frac{2\sigma}{R_o} +
 (b^2 + 1)\left(\frac{\sigma}{R_o}\right)^2},
\end{eqnarray}
where $R_o (= r_o\sqrt{b^2+1})$ is the radius of the curvature of the outer wall.
Then, the dimensionless tilt of the pressure can be rewritten as
\begin{eqnarray}
\Delta P_l^{\mathrm{BM}} = \frac{\phi_l(r_o,\theta) - \phi_l(r_o g_c,\theta)}
 {\phi_l(r_o,\theta_{\mathrm{base}})},
\end{eqnarray}
where $g_c = g\left(\frac{\sigma_c}{R_o}\right)$ and
$\sigma_c$ is the half width of the BM.
For the first and the second modes, we have
\begin{eqnarray}
\Delta P_1^{\mathrm{BM}} = -\frac{b}{2K\pi(b^2+1)}\ln g_c, \label{pp1}
\end{eqnarray}
\begin{eqnarray}
\Delta P_2^{\mathrm{BM}} = \frac{1-b^2}{(2K\pi)^2(b^2+1)^2}
 \left[\ln^2 g_c^2 - \frac{2b(b^2+1)}{1-b^2}\theta \ln g_c\right].
\label{pp2}
\end{eqnarray}

\section{Results and Discussions}
Under the long-wave approximation, $\mu^2 \ll
1$~\cite{PRL_v96_088701,JFM_v106_149}, we have obtained the
solutions of the three dimensional mechanical model of the spiral
shaped cochlea.  Now, we show that $\mu$ can be sufficiently small for any
frequency as long as the stimulation is weak. As an example, we
study the case of mode II($l=2$). Consider the velocity of
the BM near $\zeta=0$ at which the velocity of the BM reaches the
maximum. The velocity of the BM is assumed to be the same as the
velocity of the fluid on the BM. Therefore, the velocity of the BM can be
represented by the velocity of the fluid as
\begin{eqnarray}
U_{2,\zeta}(r,\theta,0,t) &=& -\mu^2 h  \phi(r,\theta).
\end{eqnarray}
In a weak stimulation, the velocity of the BM is very
small, $\sim 10^{-5}\mathrm{m/s}$ or less~\cite{PNAS_v97_11744} in
the passive system. Considering that the height of the cochlea is in the order of
millimeter\cite{JASA_v65_1007},
\begin{eqnarray}
\mu^2 \phi(r,\theta) = \frac{-U_{2,\zeta}(r,\theta,0)}{h} \sim
10^{-2}/\mathrm{s}.
\end{eqnarray}
Hence, the small $\mu$ limit is consistent with not only the long-wave
approximation but also the weak stimulation.

In the case of the lowest mode $(l=1)$ in which the flow has no normal flux, 
our results show that the
dimensionless tilt of the pressure is a monotonic function of both the
distance from the apex, $\theta$, and the radius of the curvature, $R_o$.
The maximum is at the apex, as shown in Fig~\ref{mod1max}. This is
consistent with the previous theoretical result~\cite{PRL_v96_088701}. As
shown in the Fig.~\ref{mod1max}b, the dimensionless tilt of the pressure is a
scaling function of the radius of the curvature, $\Delta P_1^{\mathrm{BM}} \sim
R_o^{-1.08}$. This is in a close agreement to the result of the WKB-like 
approximation, $\Delta P_1^{\mathrm{BM}} \sim R_o^{-1}$~\cite{PRL_v96_088701}. 
In this mode, the solution can be directly extended from the outer part 
to the inner part. According to Eqs.~(\ref{po1}) and (\ref{pressure}), the
pressure difference on the BM is also a monotonic function from the inner
wall to the outer wall. In addition to this lowest mode, we find that there
exist higher modes which show more complex behaviors.
For example, in the second ($l=2$) and the third modes ($l=3$), the positions
of the maximum of the tilt are determined by the compactness of the cochlea
(described by $b$) and the reduced width of the cochlea ($\sigma_c/\rho_o$)
as shown in Figs.~\ref{tilt}-\ref{mod3max}. In these two cases, the
solutions of the outer part cannot be directly extended to the inner part
and the role of the RM may not be neglected.
Our calculation reveals that except in the lowest mode,
the distribution of the resultant pressure acting on the BM is not only
determined by the curvature of the cochlea but also influenced by the width
of the cochlear duct (Fig.~\ref{max}), which again implies the importance
of the RM geometry in a more quantitative calculation. 
The amplitude of the tilt of the
first responsive mode is larger than the tilt of the second and the third
responsive modes in the apical region (Fig.~\ref{comp}).

The resultant pressure gradient in the radial direction induces the
vibration of the hair cells on the BM. The pattern of the pressure tilt
along the longitudinal length of the BM is therefore important for
understanding the hearing mechanism. Previous work with the WKB-like
approximation has shown that the curvature influences the tilt
monotonically~\cite{PRL_v96_088701}. However, our results reveal
that the influence of the shape can be more complex than that obtained
by Manoussaki {\it et al.}~\cite{PRL_v96_088701}.

The general solution of the pressure distribution
in the fluid will be a linear superposition of various modes.
To show the diversity of the pressure tilt along the BM, we now consider,
for simplicity, only three lowest
$\Phi_{sup} = (\alpha_1\Phi_1+\alpha_2\Phi_2 + \alpha_3\Phi_3)/(\alpha_1 +
\alpha_2 + \alpha_3)$, where
$\alpha_1$, $\alpha_2$ and $\alpha_3$ are any real constants.
In order to determine the relation between different $A_l$, we assume that
the pressure at the basal end is same to the input pressure when each
individual mode is allowed to exist alone. Since
the coefficients appearing in the linear combination for a real
oscillation cannot be determined a priori, thus, we assume two simple
ratios of $\alpha_1 : \alpha_2 : \alpha_3 = 1:2:1$ and $\alpha_1 :
\alpha_2 : \alpha_3 = 2:2:1$ and plot the tilt of the pressure
of these two examples as superpositions of the modes
in Fig.~\ref{linear}. The result shows that the maximum may not occur at
the apex, but at a position not far from the apex. We believe that the exact
maximum position is determined by the spiral and duct geometry 
of the cochlea and
the detailed nature of the pressure wave. This result clearly indicates
that the spiral curvature plays a decisive role in the wave energy
redistribution in the cochlea as reported in the earlier calculation,
although the detailed nature of the maximum position may be
different from the WKB-like approximation~\cite{PRL_v96_088701}. Here, we
note that we have assumed that the channels are fully filled with the
fluid. If the channels are partially filled, the formalae and te results
will be totally different.

\section{Conclusion}
In a long-wave approximation, the flow field of the three dimensional
fluid is obtained analytically for the spiral shaped cochlear model.
Pressure of the fluid is then calculated through the N-S equation.
Result shows that the pressure acting on the BM is not uniform along
the spanwise direction of the BM. Furthermore, the
patterns of the pressure tilts along the longitudinal length of the BM
are different for different modes. In general, the real pressure
distribution is expected to be a linear superposition of various
modes, although the simplest one may be dominant or at least substantial.
Contributions from the higher modes are expected to push the pressure
tilt maximum towards the basal end from the apex to a position which is
determined by the
curvature and the duct geometry. In conclusion, we have shown that
the curvature of the cochlea plays a decisive role in hearing in
addition to providing a compact storage.

\begin{acknowledgements}
This work was supported by the Korea Science and Engineering Foundation
(KOSEF) grant funded by the Korea government (MOST) (R01-2006-000-10083-0).
\end{acknowledgements}

\appendix
\section{Unit vectors of logarithmic spiral}

Let the position vector of the curve is $\mathbf{r}$. The unit
tangent and normal vectors, as shown in Fig.~\ref{unit}, are respectively
\begin{eqnarray}
 \hat{e}_t =
\frac{d\mathbf{r}}{d\theta}\bigg/\left|\frac{d\mathbf{r}}{d\theta}\right|,
\hat{e}_n = \frac{d\hat{e}_t}{d\theta}.
\end{eqnarray}
In the polar coordinate system, the position vector of the logarithmic spiral is
\begin{eqnarray}
\mathbf{r} = \rho \mathrm{e}^{b\theta}\hat{e}_r,
\end{eqnarray}
where $\hat{e}_r$ is the unit vector in the $r$ direction in the polar
coordinate system. Consequently,
\begin{eqnarray}
\hat{e}_t &=& \frac{1}{\sqrt{b^2+1}}\left(b\hat{e}_{r} +
\hat{e}_{\theta}\right), \label{AA_11} \\
\hat{e}_n &=& \frac{1}{\sqrt{b^2+1}}\left(-\hat{e}_{r} +
b\hat{e}_{\theta} \right). \label{AA_12}
\end{eqnarray}


\begin{thebibliography}{19}
\expandafter\ifx\csname natexlab\endcsname\relax\def\natexlab#1{#1}\fi
\expandafter\ifx\csname bibnamefont\endcsname\relax
  \def\bibnamefont#1{#1}\fi
\expandafter\ifx\csname bibfnamefont\endcsname\relax
  \def\bibfnamefont#1{#1}\fi
\expandafter\ifx\csname citenamefont\endcsname\relax
  \def\citenamefont#1{#1}\fi
\expandafter\ifx\csname url\endcsname\relax
  \def\url#1{\texttt{#1}}\fi
\expandafter\ifx\csname urlprefix\endcsname\relax\def\urlprefix{URL }\fi
\providecommand{\bibinfo}[2]{#2}
\providecommand{\eprint}[2][]{\url{#2}}

\bibitem[{\citenamefont{Flecher}(1940)}]{RMP_v12_47}
\bibinfo{author}{\bibfnamefont{H.}~\bibnamefont{Flecher}},
  \bibinfo{journal}{Rev. Mod. Phys.} \textbf{\bibinfo{volume}{12}},
  \bibinfo{pages}{47} (\bibinfo{year}{1940}).

\bibitem[{\citenamefont{Robles and Ruggero}(2001)}]{PhysiolRev_v81_1305}
\bibinfo{author}{\bibfnamefont{L.}~\bibnamefont{Robles}} \bibnamefont{and}
  \bibinfo{author}{\bibfnamefont{M.~A.} \bibnamefont{Ruggero}},
  \bibinfo{journal}{Physiol. Rev.} \textbf{\bibinfo{volume}{81}},
  \bibinfo{pages}{1305} (\bibinfo{year}{2001}).

\bibitem[{\citenamefont{Helmholtz}(1954)}]{helmholtz}
\bibinfo{author}{\bibfnamefont{H.~L.} \bibnamefont{Helmholtz}},
  \emph{\bibinfo{title}{On the Senstations of Tone as a Physiological Basis for
  the Theory of Music}} (\bibinfo{publisher}{Dover Publications Inc.},
  \bibinfo{address}{New York}, \bibinfo{year}{1954}).

\bibitem[{\citenamefont{Dallos et~al.}(1996)\citenamefont{Dallos, Popper, and
  Fay}}]{Dallos}
\bibinfo{editor}{\bibfnamefont{P.}~\bibnamefont{Dallos}},
  \bibinfo{editor}{\bibfnamefont{A.~N.} \bibnamefont{Popper}},
  \bibnamefont{and} \bibinfo{editor}{\bibfnamefont{R.~R.} \bibnamefont{Fay}},
  eds., \emph{\bibinfo{title}{The Cochlea}} (\bibinfo{publisher}{Springer},
  \bibinfo{address}{New York}, \bibinfo{year}{1996}).

\bibitem[{\citenamefont{Gummer}(2003)}]{Gummer}
\bibinfo{editor}{\bibfnamefont{A.~W.} \bibnamefont{Gummer}}, ed.,
  \emph{\bibinfo{title}{Biophysics of the Cochlea}} (\bibinfo{publisher}{World
  Scientific}, \bibinfo{address}{Singapore}, \bibinfo{year}{2003}).

\bibitem[{\citenamefont{Eatock et~al.}(2006)\citenamefont{Eatock, Fay, and
  Popper}}]{HC}
\bibinfo{editor}{\bibfnamefont{R.~A.} \bibnamefont{Eatock}},
  \bibinfo{editor}{\bibfnamefont{R.~R.} \bibnamefont{Fay}}, \bibnamefont{and}
  \bibinfo{editor}{\bibfnamefont{A.~N.} \bibnamefont{Popper}}, eds.,
  \emph{\bibinfo{title}{Vertebrate Hair Cells}} (\bibinfo{publisher}{Springer},
  \bibinfo{address}{New York}, \bibinfo{year}{2006}).

\bibitem[{\citenamefont{Hudspeth}(1997)}]{CON_v7_480}
\bibinfo{author}{\bibfnamefont{A.~J.} \bibnamefont{Hudspeth}},
  \bibinfo{journal}{Curr. Opin. Neurobio.} \textbf{\bibinfo{volume}{7}},
  \bibinfo{pages}{480} (\bibinfo{year}{1997}).

\bibitem[{\citenamefont{von B{\'{e}}k{\'{e}}sy}(1960)}]{von}
\bibinfo{author}{\bibfnamefont{G.}~\bibnamefont{von B{\'{e}}k{\'{e}}sy}},
  \emph{\bibinfo{title}{Experiments in Hearing}}
  (\bibinfo{publisher}{McGraw-Hill Book Co.}, \bibinfo{address}{New Yourk},
  \bibinfo{year}{1960}).

\bibitem[{\citenamefont{Viergever}(1978)}]{JASA_v64_1048}
\bibinfo{author}{\bibfnamefont{M.~A.} \bibnamefont{Viergever}},
  \bibinfo{journal}{J. Acoust. Soc. Am.} \textbf{\bibinfo{volume}{64}},
  \bibinfo{pages}{1048} (\bibinfo{year}{1978}).

\bibitem[{\citenamefont{Loh}(1983)}]{JASA_v74_95}
\bibinfo{author}{\bibfnamefont{C.~H.} \bibnamefont{Loh}}, \bibinfo{journal}{J.
  Acoust. Soc. Am.} \textbf{\bibinfo{volume}{74}}, \bibinfo{pages}{95}
  (\bibinfo{year}{1983}).

\bibitem[{\citenamefont{Steele and Zais}(1985)}]{JASA_v77_1849}
\bibinfo{author}{\bibfnamefont{C.~R.} \bibnamefont{Steele}} \bibnamefont{and}
  \bibinfo{author}{\bibfnamefont{J.~G.} \bibnamefont{Zais}},
  \bibinfo{journal}{J. Acoust. Soc. Am.} \textbf{\bibinfo{volume}{77}},
  \bibinfo{pages}{1849} (\bibinfo{year}{1985}).

\bibitem[{\citenamefont{Manoussaki et~al.}(2006)\citenamefont{Manoussaki,
  Dimitriadis, and Chadwick}}]{PRL_v96_088701}
\bibinfo{author}{\bibfnamefont{D.}~\bibnamefont{Manoussaki}},
  \bibinfo{author}{\bibfnamefont{E.~K.} \bibnamefont{Dimitriadis}},
  \bibnamefont{and} \bibinfo{author}{\bibfnamefont{R.~S.}
  \bibnamefont{Chadwick}}, \bibinfo{journal}{Phys. Rev. Lett}
  \textbf{\bibinfo{volume}{96}}, \bibinfo{pages}{088701}
  (\bibinfo{year}{2006}).

\bibitem[{\citenamefont{Kern}(2003)}]{KernThesis}
\bibinfo{author}{\bibfnamefont{A.}~\bibnamefont{Kern}}, Ph.D. thesis,
  \bibinfo{school}{Swiss Federal Institute of Technology (ETH)},
  \bibinfo{address}{Z{\"{u}}rich} (\bibinfo{year}{2003}).

\bibitem[{\citenamefont{Homer et~al.}(2004)\citenamefont{Homer, Champneys,
  Hunt, and Cooper}}]{JASA_v116_1025}
\bibinfo{author}{\bibfnamefont{M.}~\bibnamefont{Homer}},
  \bibinfo{author}{\bibfnamefont{A.}~\bibnamefont{Champneys}},
  \bibinfo{author}{\bibfnamefont{G.}~\bibnamefont{Hunt}}, \bibnamefont{and}
  \bibinfo{author}{\bibfnamefont{N.}~\bibnamefont{Cooper}},
  \bibinfo{journal}{J. Acoust. Soc. Am} \textbf{\bibinfo{volume}{116}},
  \bibinfo{pages}{1025} (\bibinfo{year}{2004}).

\bibitem[{\citenamefont{Lighthill}(1981)}]{JFM_v106_149}
\bibinfo{author}{\bibfnamefont{J.}~\bibnamefont{Lighthill}},
  \bibinfo{journal}{J. Fluid Mech.} \textbf{\bibinfo{volume}{106}},
  \bibinfo{pages}{149} (\bibinfo{year}{1981}).

\bibitem[{\citenamefont{Riley et~al.}(2006)\citenamefont{Riley, Hobson, and
  Bence}}]{math}
\bibinfo{author}{\bibfnamefont{K.~F.} \bibnamefont{Riley}},
  \bibinfo{author}{\bibfnamefont{M.~P.} \bibnamefont{Hobson}},
  \bibnamefont{and} \bibinfo{author}{\bibfnamefont{S.~J.} \bibnamefont{Bence}},
  \emph{\bibinfo{title}{Mathematical Methods for Physics and Engineering}}
  (\bibinfo{publisher}{Cambridge University Press},
  \bibinfo{address}{Cambridge}, \bibinfo{year}{2006}), \bibinfo{edition}{3rd}
  ed.

\bibitem[{\citenamefont{Lighthill}(2005)}]{Lighthill}
\bibinfo{author}{\bibfnamefont{J.}~\bibnamefont{Lighthill}},
  \emph{\bibinfo{title}{Waves in Fluid}} (\bibinfo{publisher}{Cambridge
  University press}, \bibinfo{address}{Cambridge}, \bibinfo{year}{2005}).

\bibitem[{\citenamefont{Ruggero et~al.}(2000)\citenamefont{Ruggero, Narayan,
  Temchin, and Recio}}]{PNAS_v97_11744}
\bibinfo{author}{\bibfnamefont{M.~A.} \bibnamefont{Ruggero}},
  \bibinfo{author}{\bibfnamefont{S.~S.} \bibnamefont{Narayan}},
  \bibinfo{author}{\bibfnamefont{A.~N.} \bibnamefont{Temchin}},
  \bibnamefont{and} \bibinfo{author}{\bibfnamefont{A.}~\bibnamefont{Recio}},
  \bibinfo{journal}{Proc. Natl. Acd. Sci. USA} \textbf{\bibinfo{volume}{97}},
  \bibinfo{pages}{11744} (\bibinfo{year}{2000}).

\bibitem[{\citenamefont{Steele and Taber}(1979)}]{JASA_v65_1007}
\bibinfo{author}{\bibfnamefont{C.~R.} \bibnamefont{Steele}} \bibnamefont{and}
  \bibinfo{author}{\bibfnamefont{L.~A.} \bibnamefont{Taber}},
  \bibinfo{journal}{J. Acoust. Soc. Am.} \textbf{\bibinfo{volume}{65}},
  \bibinfo{pages}{1007} (\bibinfo{year}{1979}).

\end{thebibliography}

\newpage
\begin{figure}[htbp]
\begin{center}
\includegraphics*[width=0.7\columnwidth]{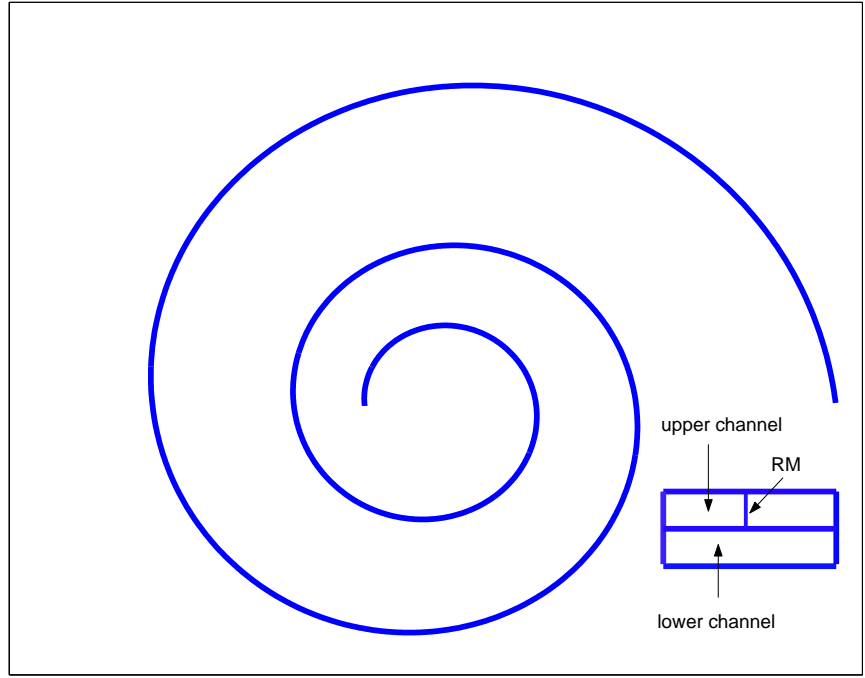}
\end{center}
\caption{Sketch of the shape of the cochlea in the present model. The top view of
the cochlea is modeled as a logarithmic spiral; the cross section of
the cochlea is approximated as a rectangle with constant width and
height along the entire length of the cochlea. The duct of the
cochlea is divided by the BM into two channels.} \label{model}
\end{figure}

\begin{figure}[htbp]
\begin{center}
\includegraphics*[width=0.4\columnwidth]{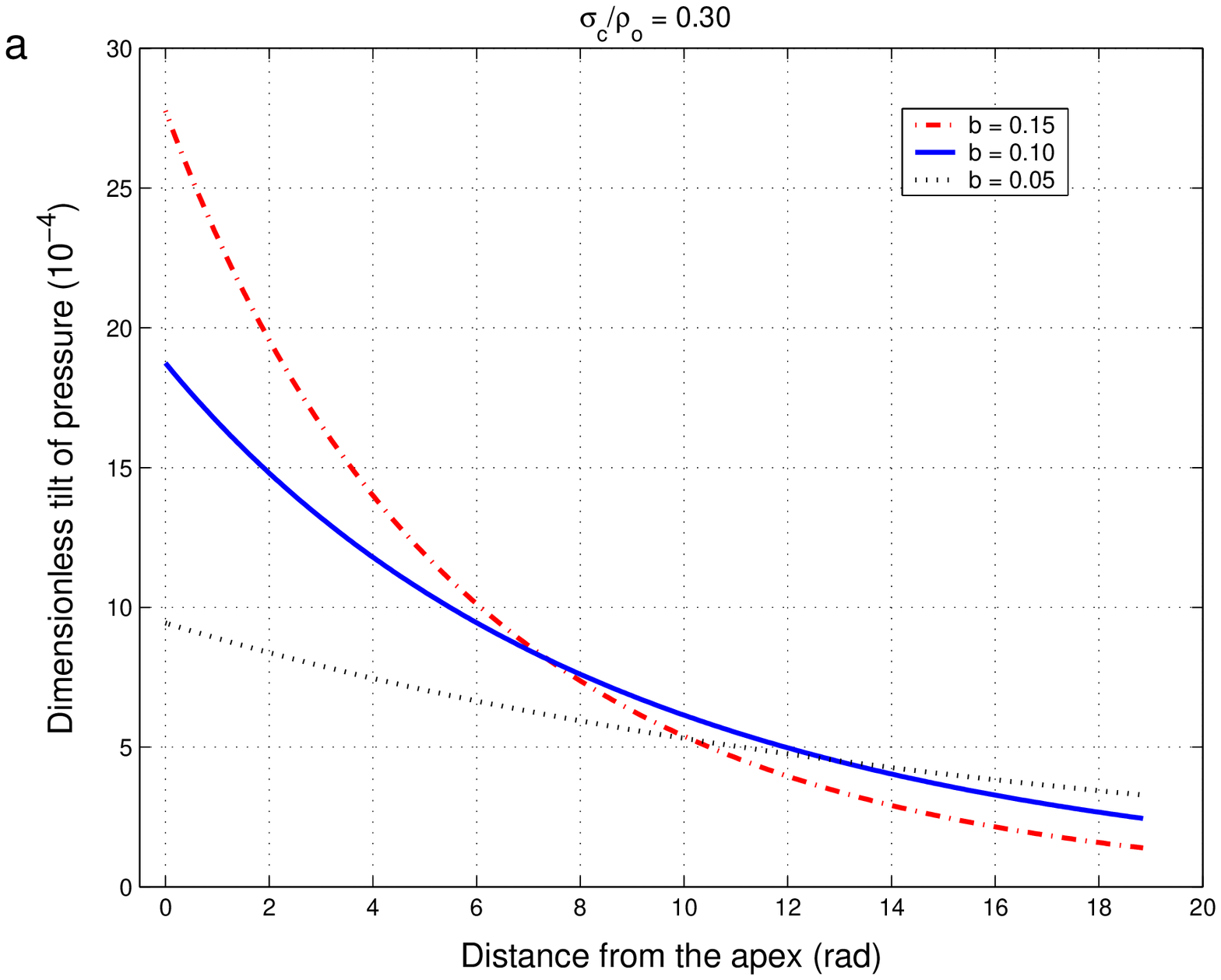}
\includegraphics*[width=0.4\columnwidth]{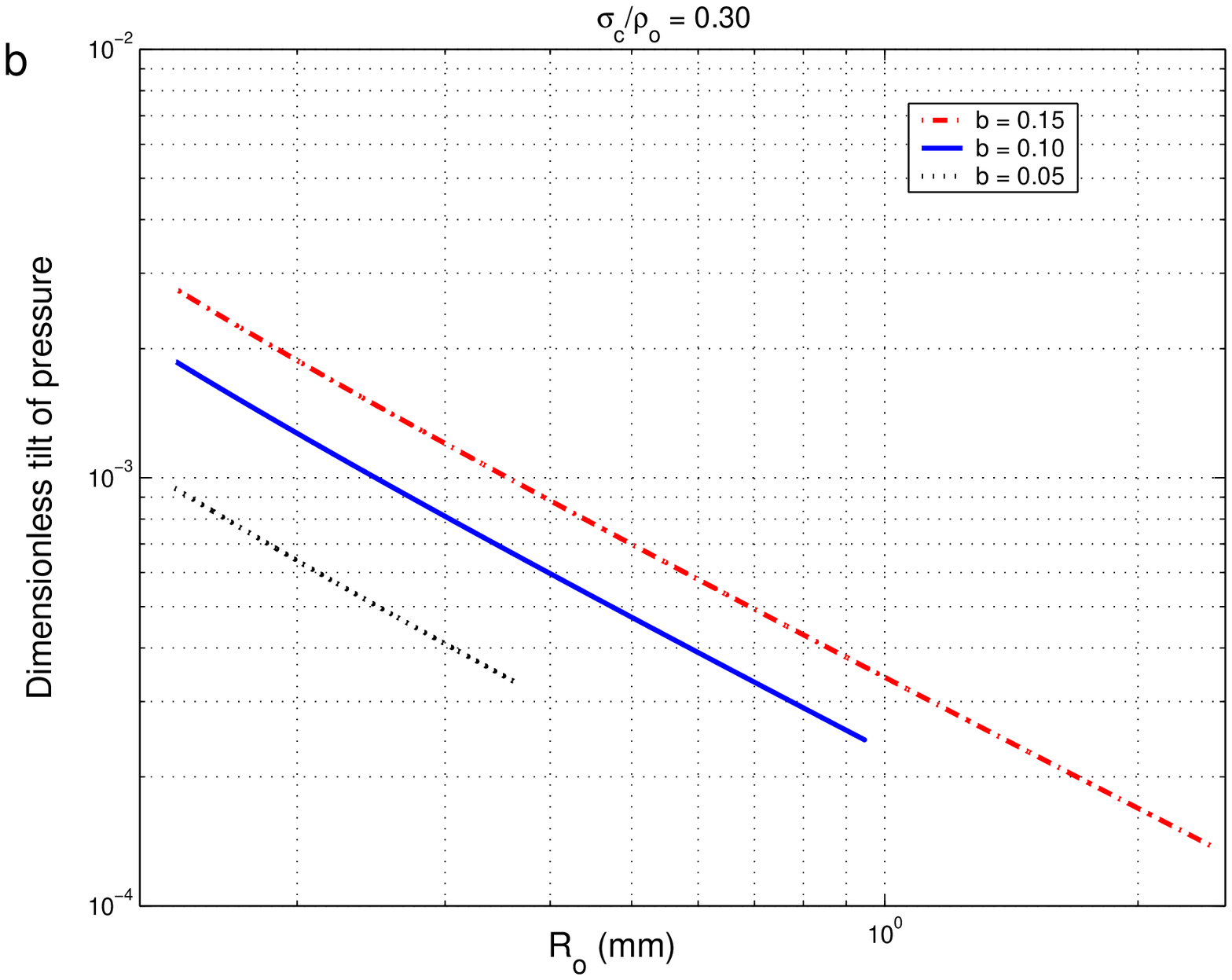}
\end{center}
\caption{(color online). The pressure tilt in the mode I ($l=1$).
The maximum always appears at the apical end. For $\sigma_c/\rho_o$, 
we use $0.3$. (a): The tilt is plotted as a function of the 
distance from the apex. (b): The dimensionless pressure tilt 
is a scaling function of the radius of the curvature, $R_o$.
The slope of the straight lines is about $1.08$.
Therefore, it reveals that, in this mode, our result is consistent with the
previous theoretical result~\cite{PRL_v96_088701}.}
\label{mod1max}
\end{figure}

\begin{figure}[htbp]
\begin{center}
\includegraphics*[width=0.7\columnwidth]{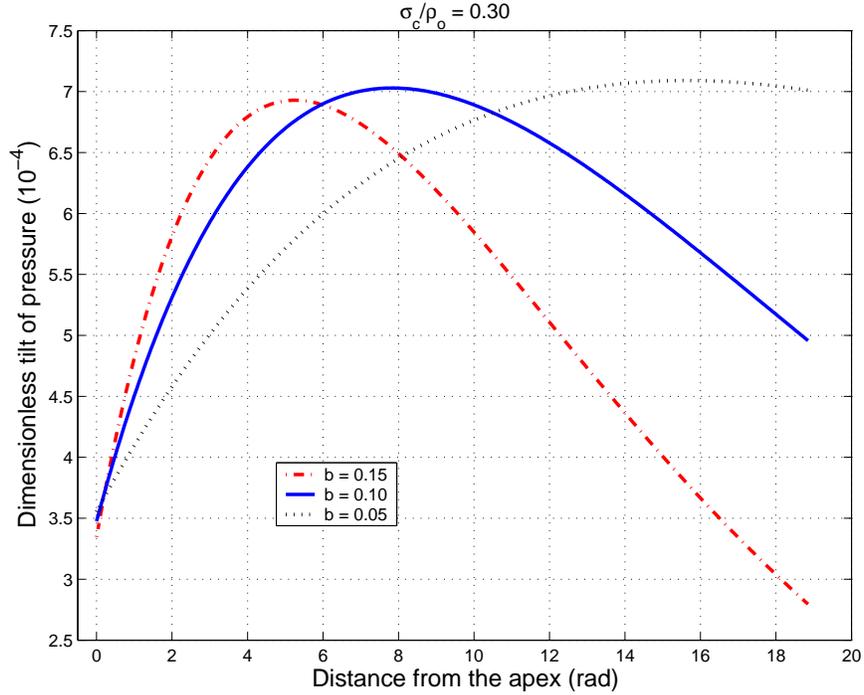}
\end{center}
\caption{(color online). In the mode II ($l=2$),
the tilt of the resultant pressure
acting on the BM. It reaches a maximum and decreases along the
longitudinal length of the cochlea. The position of the maximum
depends on $\sigma_c/\rho_o$ and $b$. } \label{tilt}
\end{figure}

\begin{figure}[htpb]
\begin{center}
\includegraphics*[width=0.7\columnwidth]{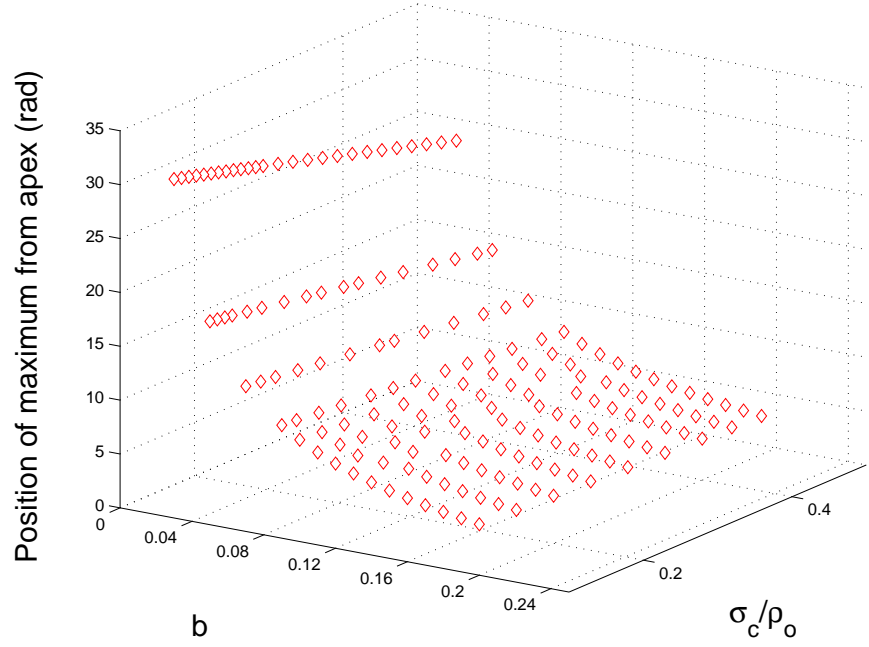}
\end{center}
\caption{In the mode II, the maximum of the tilt appears at a
position along the longitudinal length of the cochlea. It is
determined by both the compactness and the width of the cochlea.
}\label{max}
\end{figure}

\begin{figure}[htbp]
\begin{center}
\includegraphics*[width=0.7\columnwidth]{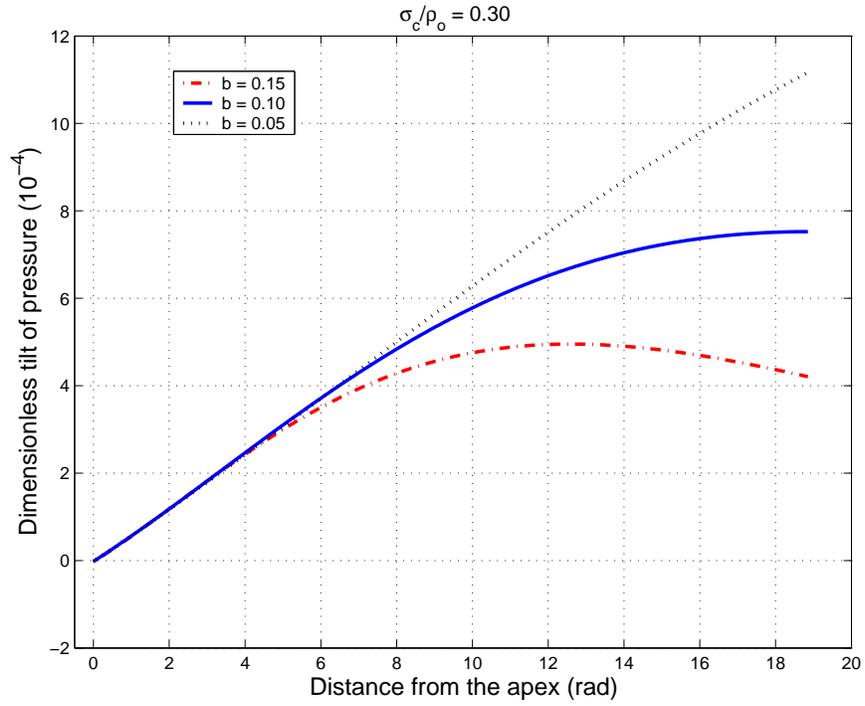}
\end{center}
\caption{(color online). The tilt of the pressure in the mode III
($l = 3$). The curves are calculated at $\sigma_c/\rho_o = 0.3$.
The maximum of the tilt appears at
a position determined by the geometry of the cochlea.}\label{mod3max}
\end{figure}

\begin{figure}[htbp]
\begin{center}
\includegraphics*[width=0.7\columnwidth]{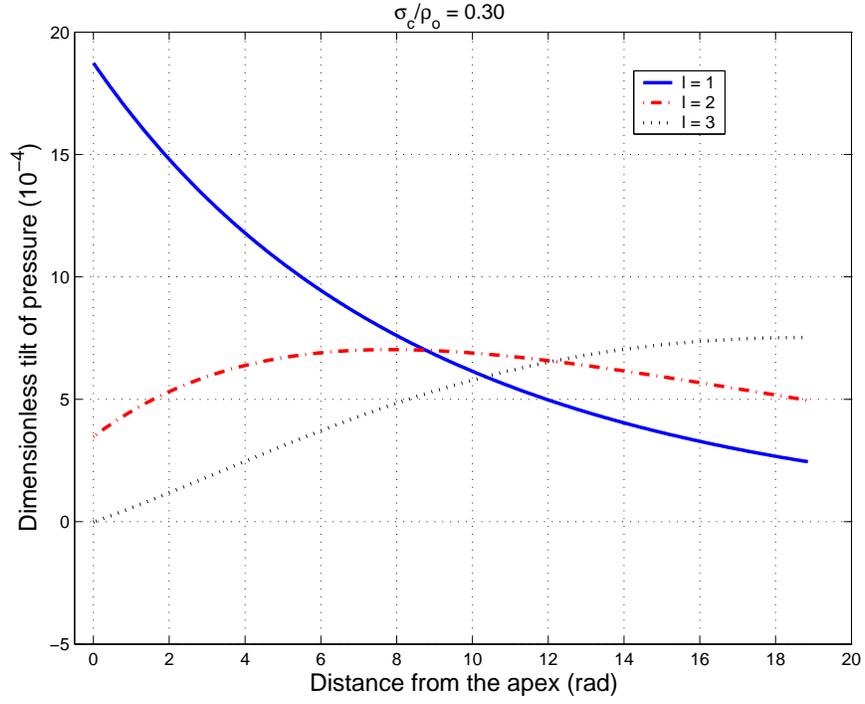}
\end{center}
\caption{ (color online).
A comparison of the tilt of the resultant pressure for the first
three modes. $b=0.10$, $K=3$ and $\sigma_c/\rho_o=0.3$. In the
apical region, the tilt of the first responsive mode is larger than
the tilt of the second mode; The tilt of the second responsive mode
is larger than the tilt of the third mode.}
\label{comp}
\end{figure}

\begin{figure}[htbp]
\begin{center}
\includegraphics*[width=0.7\columnwidth]{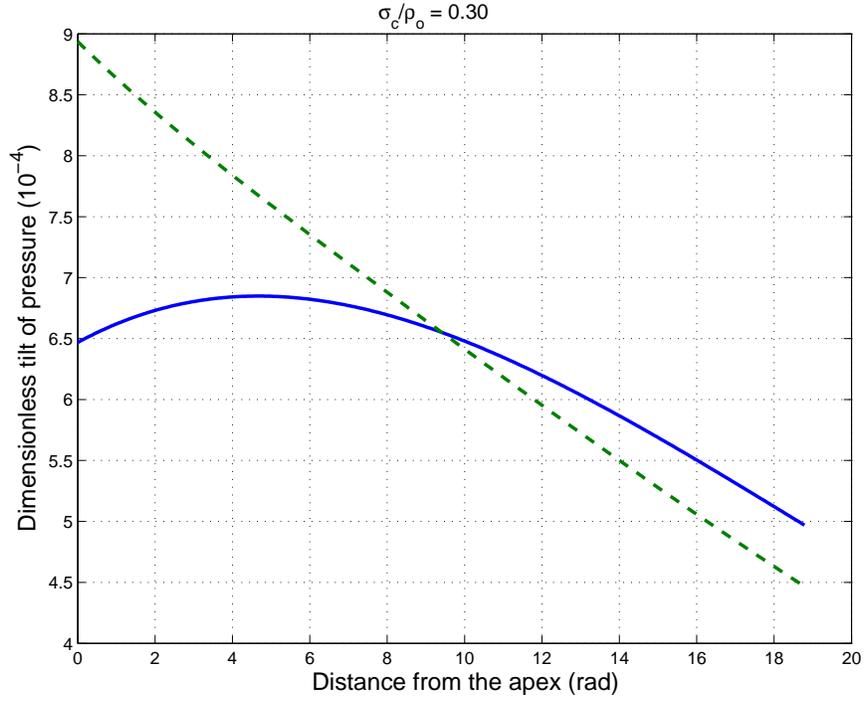}
\end{center}
\caption{(color online).
Linear superpositions of the first three modes. The maximum
of the tilt is located at a position determined by the coefficients
of the superposed modes and the geometry of the cochlea. $b = 0.10$,
$K=3$ and $\sigma_c/\rho_o = 0.30$. Solid line corresponds to
$\alpha_1 : \alpha_2 : \alpha_3 =1:2:1$. Dash line corresponds to
$\alpha_1 : \alpha_2 : \alpha_3 = 2:2:1$. }\label{linear}
\end{figure}

\begin{figure}[htbp]
\begin{center}
\includegraphics*[width=0.7\columnwidth]{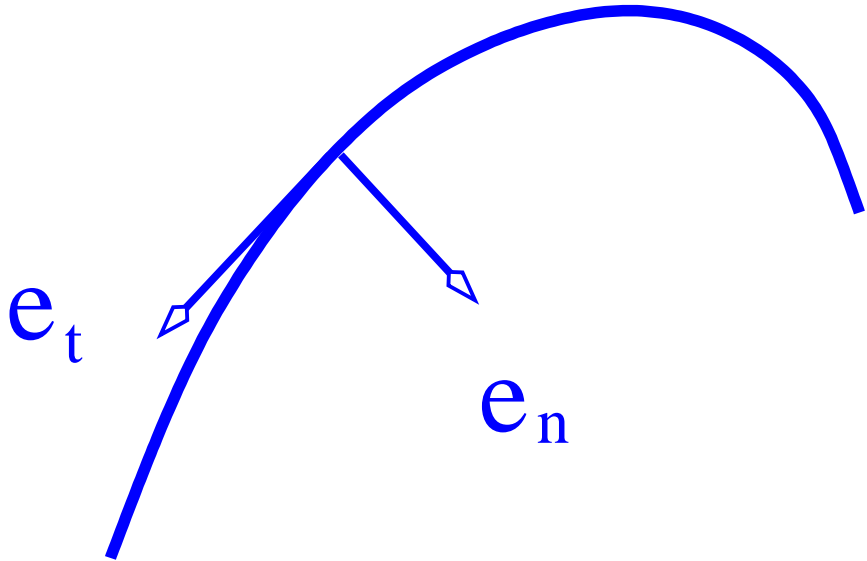}
\end{center}
\caption{The unit tangent vector, $\hat{e}_t$, and the unit normal vector,
$\hat{e}_n$, of a logarithmic spiral. }\label{unit}
\end{figure}

\end{document}